\documentclass[11pt,a4paper]{article}
\usepackage{jheppub}
\usepackage[utf8]{inputenc}
\usepackage{amsmath}
\usepackage{amsfonts}
\usepackage{amssymb}
\usepackage{graphicx}
\usepackage{caption}
\usepackage{subcaption}
\usepackage{float}
\pdfoutput=1
\begin{document}
\title{Time dependence of entanglement entropy on the fuzzy sphere} 
\author[a]{Philippe Sabella-Garnier}
\affiliation[a]{Lorentz Institute for Theoretical Physics, Leiden University \\ Niels Bohrweg 2, Leiden 2333-CA, The Netherlands}
\keywords{Non-commutative geometry, Matrix models}
\arxivnumber{1705.01969}

\emailAdd{garnier@lorentz.leidenuniv.nl}

\abstract{We numerically study the behaviour of entanglement entropy for a free scalar field on the noncommutative (``fuzzy'') sphere after a mass quench. It is known that the entanglement entropy before a quench violates the usual area law due to the non-local nature of the theory. By comparing our results to the ordinary sphere, we find results that, despite this non-locality, are compatible with entanglement being spread by ballistic propagation of entangled quasi-particles at a speed no greater than the speed of light. However, we also find that, when the pre-quench mass is much larger than the inverse of the short-distance cutoff of the fuzzy sphere (a regime with no commutative analogue), the entanglement entropy spreads faster than allowed by a local model.}

\maketitle
\flushbottom
\section{Motivation and summary}
\label{intro}
The study of entanglement entropy in field theories has attracted renewed interest since it was first equated to the area of bulk minimal surfaces using AdS/CFT \cite{Ryu:2006bv}. While this original proposal only dealt with static geometries, it was soon expanded to time-dependent ones \cite{Hubeny:2007xt}. With this, it has become possible to study the time evolution of entanglement entropy after quenches. In  \cite{Liu:2013iza}, the idea of an ``entanglement tsunami'' was proposed:\footnote{See also \cite{Calabrese:2005in} for a pre-holographic approach.} after a global quench, entanglement is generated extensively and soon\footnote{This tsunami picture is only valid after a time set by the inverse effective temperature of the quench. If the subsystem size is small enough, this model may fail \cite{Kundu:2016cgh}.}  spreads balistically. This results in the entanglement entropy of a region growing linearly for some time after the quench and saturating after a time equal to the linear size of the region. This picture can be refined and better understood in terms of a quasi-particle model where all the entanglement is carried by EPR pairs created during the quench and moving at a speed bounded above by the speed of light \cite{Casini:2015zua}. The entanglement entropy in a region $\Sigma$ of characteristic linear size $R$ is then expected to behave as
\begin{equation}
S(t,R)=  \begin{cases}
\mathfrak{s}_\Sigma(R) F_\Sigma(t) +S_0 ~,~ t \leq R \\
\mathfrak{s}_\Sigma(R) +S_0 ~,~ t>R ~,
\end{cases} 
\label{model}
\end{equation}
where  $F_\Sigma(t)$ is a function depending on the geometry of $\Sigma$ with the property that $F_\Sigma(0)=0$ and $F_\Sigma(R)=1$, $S_0$ is the (divergent) initial entanglement entropy and $\mathfrak{s}_\Sigma(R)$ sets the overall scale. If the geometry of $\Sigma$ is relatively simple, analytic expressions can be obtained for the above expressions. For example, in \cite{Casini:2015zua} the entanglement entropy of a disk of radius $R$ in two dimensions was predicted to follow
\begin{align}
F_{\Sigma}(x) &=\frac{2}{\pi} \left(x\sqrt{1-x^2} + \arcsin x \right) ~,~ x=t/R~, \label{flat} \\
\mathfrak{s}_{\Sigma}(R)&=s \pi R^2~.
\end{align} 
In this case, $\mathfrak{s}(R)$ scales as the area of the disk. More generally, $\mathfrak{s}_\Sigma(R)$ is expected to be extensive since the model assumes a uniform creation of entangled pairs. $s$ can then be thought of as the density of entanglement pairs produced by the quench. This model was found to be in good agreement with numerical calculations for a global quench in a free field theory \cite{Cotler:2016acd}. In that paper, it was also shown that, for a mass quench (taking the free field mass to go from $M$ to $0$), $s$ follows:
\begin{align}
s=\frac{\log 2}{4\pi} M^2~.
\label{prop}
\end{align}

In this note, we examine numerically how well \ref{model} describes the behaviour of the entanglement entropy of a free scalar field on a noncommutative sphere after a mass quench. We take the region $\Sigma$ to be a polar cap of varying angular size (the angular size playing the role of $R$). This setup is the easiest non-trivial one for a noncommutative theory.\footnote{Noncommutativity has no effect on free field theory on a plane} Nonetheless, the geometrical problem of finding $F_\Sigma(t)$ is rather intractable analytically, therefore we use a numerical calculation of the same setup in the commutative theory as a proxy for the details of the quasi-particle model. This is reasonable given that the quasi-particle model has been found to describe the aftermath of a mass quench in a free field theory in flat space well \cite{Cotler:2016acd} and that the general features should not change when we change the geometry.

In addition to being related to the behaviour of D-branes in magnetic fields \cite{Seiberg:1999vs} and the quantum Hall effect \cite{Bigatti:1999iz}, theories defined on noncommutative geometries are interesting because they are inherently non-local. Non-locality has been postulated to be essential to fast scrambling \cite{Sekino:2008he, Lashkari:2011yi, Brady:2013opa}. Entanglement entropy on the fuzzy sphere has been shown to deviate from the area law even in the limit where one would expect noncommutativity to vanish \cite{Karczmarek:2013jca,Sabella-Garnier:2014fda,Okuno:2015kuc,Suzuki:2016sca}.\footnote{Other studies of field theory in non-local theories have also seen violation of the area law for entanglement entropy, see for example \cite{Fischler:2013gsa,Karczmarek:2013xxa,Shiba:2013jja,Pang:2014tpa,MohammadiMozaffar:2017nri}.} 

The rest of this note is organized as follows. In Section \ref{setup}, we explain the exact setup for these calculations: we review some relevant concepts about the noncommutative sphere, present the Hamiltonian for each of the two theories we focus on and summarize how we calculate entanglement entropy after a mass quench. We present our results in Section \ref{results}: we compare the commutative and noncommtutative theories, examining both $F_\Sigma(t)$ and $\mathfrak{s}_\Sigma(R)$. We find that for masses well below the UV cutoff there is no sign of non-locality in the time dependence, area dependence or mass dependence of the entanglement entropy following a quench. This is our main result. However, we note that if we consider a mass larger than the UV cutoff imposed by the noncommutativity of the fuzzy sphere, the spread of entanglement appears to be faster and its value at the expected saturation time grows more weakly as a function of mass.
We leave a more systematic study of the large mass regime as well as an analysis of the impact of coupling (either weak, as in \cite{Okuno:2015kuc,Suzuki:2016sca} or strong as in \cite{Fischler:2013gsa,Karczmarek:2013xxa}) to future work.

\section{Setup}
\label{setup}
\subsection{The fuzzy sphere}
The noncommutative (or ``fuzzy'') sphere is  obtained by replacing Cartesian coordinates $(x_1,x_2,x_3)$ with operators $(X_1,X_2,X_3)$ proportional to the $SU(2)$ generators in the irreducible representation of dimension $N=2J+1$ \cite{Madore:1991bw}:
\begin{equation}
X_i=\frac{R}{\sqrt{J(J+1)}} L_i ~,~ \left[L_i,L_j\right]=i\epsilon_{ijk} L_k ~~~~~~i=1,2,3~.
\end{equation}
Notice that these have the property that
\begin{equation}
X_1^2+X_2^2+X_3^2=R^2 \bold{1}_N~,
\end{equation}
where $\bold{1}_N$ is the $N\times N$ identity matrix, whence the spherical geometry. Evaluating any function at a particular point on the sphere is done by taking the expectation value of the equivalent operator in a corresponding coherent state (as reviewed in \cite{Karczmarek:2013jca}). These states are necessarily overcomplete, which means that the overlap between states corresponding to different points is not zero. In fact, the width of a state is proportional to a certain noncommutativity lengthscale $\frac{R}{\sqrt{N}}$. The number of coherent states is $N^2$, thus we have a small-distance (UV) cutoff proportional to $\frac{R}{N}$. Integration is accomplished by taking a trace and differentiation by taking the commutator with coordinate operators.\footnote{This can be made more precise by considering fuzzy spherical harmonics, obtained by replacing Cartesian coordinates with their matrix analogues in the expressions for spherical coordinates. For a review of noncommutative geometry, see e.g. \cite{Douglas:2001ba}.}
\subsection{Free scalar field on the fuzzy sphere}
\begin{figure}
\centering
\includegraphics[scale=0.3]{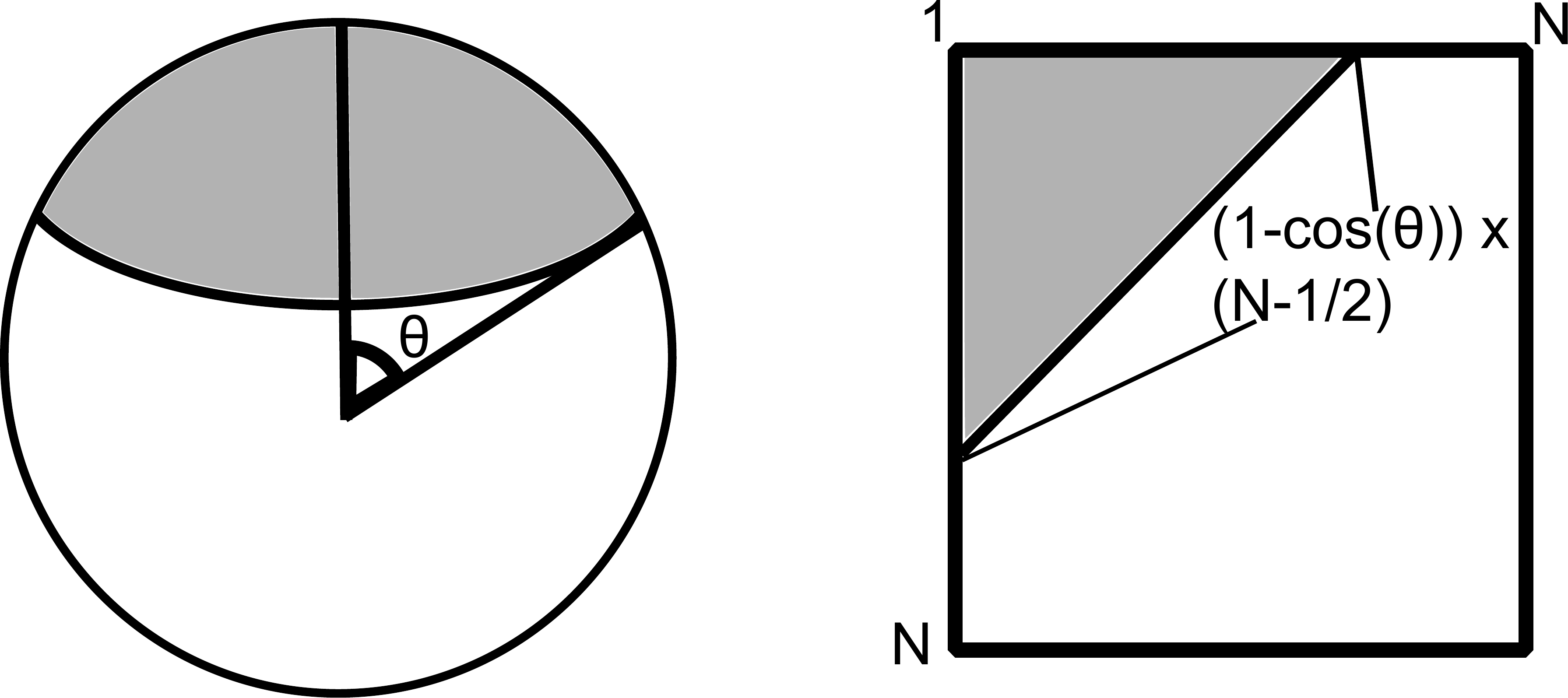}
\caption{Degrees of freedom on a polar cap and the corresponding matrix elements. Figure taken from \cite{Sabella-Garnier:2014fda}.}
\label{dof}
\end{figure}

The Hamiltonian of free field theory on a noncommutative sphere takes the form
\begin{equation}
H=\frac{2\pi R^2}{N} \text{Tr}\left(\Pi^2-R^{-2}\left[L_i,\Phi\right]^2+\mu^2 \Phi^2\right)~,
\end{equation}
where $\Phi$ is the scalar field (an $N\times N$ matrix), $\Pi$ is its conjugate momentum and $\mu$ can be thought of as the mass, coupling to the geometry, or a combination of both. From now on, we set $R=1$. This Hamiltonian is quadratic in the matrix elements of $\Phi$, therefore it can be written in the form
\begin{equation}
H=\frac{1}{2}\sum_{a,b=1}^{N^2} \left(\pi_a \delta_{ab} \pi_b + \phi_a K_{ab} \phi_b \right)~,
\end{equation}
with $\phi_a$ and $\pi_a$ being matrix elements of $\Phi$ and $\Pi$ respectively and $[\phi_a,\pi_b]=i\delta_{ab}$. In fact, due to the structure of the $L_i$, the Hamiltonian can be split into different non-interacting sectors. This decomposition, first shown in \cite{Dou:2006ni}, is as follows. Take, for $m\geq0$,
\begin{equation}
Q^{(m)}=(\Phi_{1,1+m},\Phi_{2,2+m},\dots,\Phi_{N-m,N}) ~,~ Q^{(-m)}=(\Phi_{1+m,1},\Phi_{2+m,2},\dots,\Phi_{N,N-m})~,
\end{equation}
and define the following quantities
\begin{align}
c_2=J(J+1)  \nonumber \\
A_a=-a + \frac{N+1}{2} ~~~a=1\dots N \nonumber \\
B_a=\sqrt{a(N-a)}~~~a=1\dots N,
\end{align}
where $c_2$ is the quadratic Casimir of the spin-$J$ representation of $SU(2)$ and $A_a, B_a$ are defined such that they are the non-zero elements of $L_3$ and $L_{\pm}$ respectively. Then the Hamiltonian can be written as
\begin{align}
H&= \sum_{m=-(N-1)}^{N-1} H_m = \frac{1}{2} \sum_{m=-(N-1)}^{N-1} \sum_{a,b=1}^{N-|m|} \left( \pi_a^{(m)} \delta_{ab} \pi_b^{(m)} + Q_{a}^{(m)} V_{ab}^{(m)} Q_b^{(m)} \right)~, \\
V_{ab}^{(m)}&=\left[2\left(c_2 + \frac{\mu^2}{2}-A_a A_{a+|m|}\right)\delta_{a,b} - B_{a-1}B_{a-1+|m|}\delta_{a-1,b} - B_a B_{a+|m|}\delta_{a+1,b}\right]~.
\end{align}
In order to calculate entanglement entropies for subregion, we must know which matrix elements correspond to which regions on the sphere. In \cite{Karczmarek:2013jca}, it was shown that the matrix elements above the $k^{\text{th}}$ anti-diagonal correspond to the degrees of freedom on a polar cap of size
\begin{equation}
\cos \theta =1 - \frac{k}{N-\frac{1}{2}}~,
\end{equation}
as illustrated in figure \ref{dof}. The thickness of the boundary between the polar cap and its complement is proportional to the noncommutativity lengthscale $\frac{1}{\sqrt{N}}$.

\subsection{Free scalar field on the commutative sphere}
The Hamiltonian for the free scalar field on a commutative sphere of unit radius is
\begin{equation}
H=\frac{1}{2}\int d\Omega \left(\Pi^2 + \left(\frac{\partial \Phi}{\partial \theta}\right)^2 + \frac{1}{\sin^2\theta} \left(\frac{\partial \Phi}{\partial \phi}\right)^2 + \mu^2 \Phi^2\right)~.
\end{equation}
To calculate entanglement entropy on the commutative sphere, we regularize the theory by discretizing the polar angle $\theta$:
\begin{equation}
\theta \rightarrow \theta_n = n\frac{\pi}{N}~,~~ n=1 \dots N-1~.
\end{equation}
This leads to a short-distance cutoff proportional to $\frac{1}{N}$. We then expand the field in Fourier modes along the azimuthal direction $\phi$, labelling each mode by $m$. After various rescalings necessary to ensure canonical commutation relations, we get that the Hamiltonian can be written as \cite{Sabella-Garnier:2014fda}:
\begin{equation}
H=\frac{1}{2}\sum_{m=-\infty}^\infty \sum_{i,j=1}^{N-1} \Pi^{(m)}_{i} \delta_{ij} \Pi^{(m)}_{j} + \Phi_{i}^{(m)} K_{ij}^{(m)} \Phi_j^{(m)}~,
\end{equation}
with
\begin{align}
K_{ii}^{(m)}&=\left(\frac{2N^2}{\pi^2} \cos \frac{\pi}{2N}+\frac{m^2}{\sin^2 \theta_i} + \mu^2 \right) ~~~i=2\dots N-2 ~,\nonumber\\
K_{i,i+1}^{(m)}=K_{i+1,i}^{(m)}&=-\frac{\sin \theta_{i+1/2}}{\sqrt{\sin \theta_i \sin \theta_{i+1}}} ~~~i=2\dots N-3 ~,\nonumber \\
K_{11}^{(m)}&=\frac{N^2}{2\pi^2}\frac{\sin \theta_{3/2}}{\sin\theta_1} + \frac{1}{4}\left(\frac{m^2}{\sin^2\theta_1}+\mu^2\right) ~, \nonumber\\
K_{N-1,N-1}^{(m)}&=\frac{N^2}{2\pi^2}\frac{\sin \theta_{N-3/2}}{\sin\theta_{N-1}} + \frac{1}{4}\left(\frac{m^2}{\sin^2\theta_{N-1}}+\mu^2\right)~, \nonumber\\
K_{1,2}^{(m)}=K_{2,1}^{(m)}&= - \frac{\sin \theta_{3/2}}{\sqrt{2\sin\theta_1 \sin \theta_2}} ~,\nonumber \\
K_{N-1,N-2}^{(m)}=K_{N-2,N-1}^{(m)}&= - \frac{\sin \theta_{N-3/2}}{\sqrt{2\sin\theta_{N-1} \sin \theta_{N-2}}}~.
\end{align}
\subsection{Entanglement entropy for quadratic Hamiltonians}
The Hamiltonian for both of our theories takes the form
\begin{equation}
H=\sum_{m} H_m =  \sum_{m} \sum_{i,j} \frac{1}{2} \left( p_i^{(m)} \delta_{ij} p_j^{(m)} + x^{(m)}_i K_{ij}^{(m)} x_j^{(m)} \right)~.
\end{equation}
For each $m$ sector, it is possible to write down the ground state explicitly in terms of $K^{(m)}$ and calculate the entanglement entropy \cite{Srednicki:1993im,Casini:2009sr}. Furthermore, it is possible to calculate the new ground state after a mass quench and therefore the time evolution of the entanglement entropy. This is done explicitly in \cite{Cotler:2016acd}, and we summarize that construction here. Let $O$ be the orthogonal matrix that diagonalizes $K^{(m)}$:
\begin{equation}
K^{(m)}=O^{T}K_D O~,
\end{equation}
where $K_D$ is diagonal. Denote by $\omega^2$ the vector formed by the eigenvalues of $K^{(m)}$ (i.e. the elements of $K_D$) and $\tilde{\omega}^2=\omega^2-\Delta (\mu^2)$, where $\Delta(\mu^2)$ is the difference in the squares of the pre-quench and post-quench masses. $\omega$ and $\tilde{\omega}$ are the frequencies of the normal modes of our system of coupled harmonic oscillators before and after the quench respectively.\footnote{The astute reader may notice that our definition of $\tilde{\omega}$ only works if the change in the matrix $K^{(m)}$ after the quench is proportional to the identity matrix but that the coefficients of $\mu^2$ in $K^{(m)}_{11}$ and $K^{(m)}_{N-1,N-1}$ are different than those in $K_{ii}$ for other $i$. This can be remedied by changing that coefficient, which is equivalent to making the quench slightly non-uniform. However, this is an edge effect that in practice has no impact when considering regions which are larger than about 10 sites.} Now, define the following matrices
\begin{align}
Q_{ij}=\sum_k O_{ki}O_{kj} q_k^2~, \nonumber \\
P_{ij}=\sum_k O_{ki}O_{kj} p_k^2~, \nonumber \\
R_{ij}=\sum_k O_{ki}O_{kj} r_k^2~, 
\end{align}
where 
\begin{align}
q_k^2&=\frac{1}{4\tilde{\omega}^2_k \omega_k}\left[\omega^2_k+\tilde{\omega}^2_k - (\omega^2_k-\tilde{\omega}^2_k)\cos 2\tilde{\omega}_k t\right]~,\nonumber \\
p_k^2&=\frac{1}{4\omega_k}\left[\omega^2_k+\tilde{\omega}^2_k + (\omega^2_k-\tilde{\omega}^2_k)\cos 2\tilde{\omega}_k t\right]~,\nonumber \\
r_k&=\frac{\omega^2_k-\tilde{\omega}^2_k}{4\omega_k\tilde{\omega}_k}\sin2\tilde{\omega}_k t~.
\end{align}
$q_k^2$, $p_k^2$ and $r_k$ are respectively the expectation value of $\hat{q_k}^2$, $\hat{p_k}^2$ and $\frac{1}{2}\{\hat{q_k},\hat{p_k}\}$ when the post-quench Hamiltonian is $H=\frac{1}{2} \sum_k (\hat{p_k}^2 + \tilde{\omega}_k^2 \hat{q_k}^2)$. 

Take the subsystem whose entanglement entropy we wish to calculate to be the set of oscillators $x_i$ with $i\leq I$. Define the matrix 
\begin{equation}
M=i \begin{bmatrix}
R_{ij} & P_{ij} \\
-Q_{ij} & -R_{ij}
\end{bmatrix}~,
\end{equation}
where $i,j \leq I$. This matrix has eigenvalues $\{\pm\gamma_{k}\}$ with $k=1\dots I$. The entanglement entropy is then
\begin{equation}
S^{(m)}=\sum_{k=1}^I \left[\left(\gamma_k+\frac{1}{2}\right)\log\left(\gamma_k+\frac{1}{2}\right)-\left(\gamma_k-\frac{1}{2}\right)\log\left(\gamma_k-\frac{1}{2}\right)\right]~,
\end{equation}
and the total entanglement entropy is
\begin{equation}
S=\sum_m S^{(m)}~.
\end{equation}
\section{Results}
\label{results}
With the methods explained in Section \ref{setup}, we can calculate the entanglement entropy for a scalar field theory after a mass quench on both the commutative and fuzzy spheres. For convenience, we will take $\mu=0.5$ after the quench in all cases: this corresponds to conformally coupling the  scalar to the geometry. We will mostly take the pre-mass quench to be much smaller than the UV cutoff of each theory, which in both cases means that
\begin{equation}
\mu \ll N~. \label{smallregime}
\end{equation}
Note that $\mu$ is  measured in units of the inverse sphere radius (which we have set to one), not in units of the UV cutoff; in other words as we take the continuum limit for the commutative theory we should keep $\mu$ fixed. At the end of this section, we  discuss what happens in the noncommutative theory as we take $\mu$ to be large. 

\subsection{Time dependence}
\begin{figure}[h]
\includegraphics{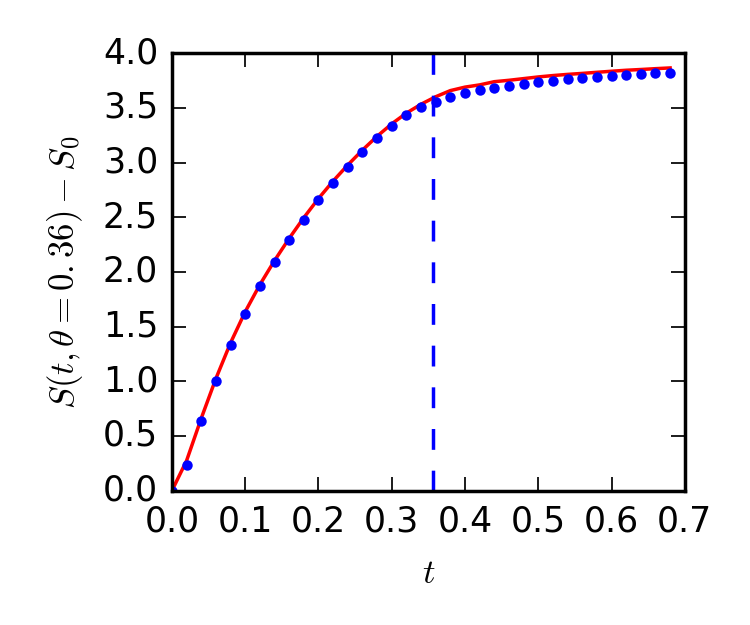}
\includegraphics{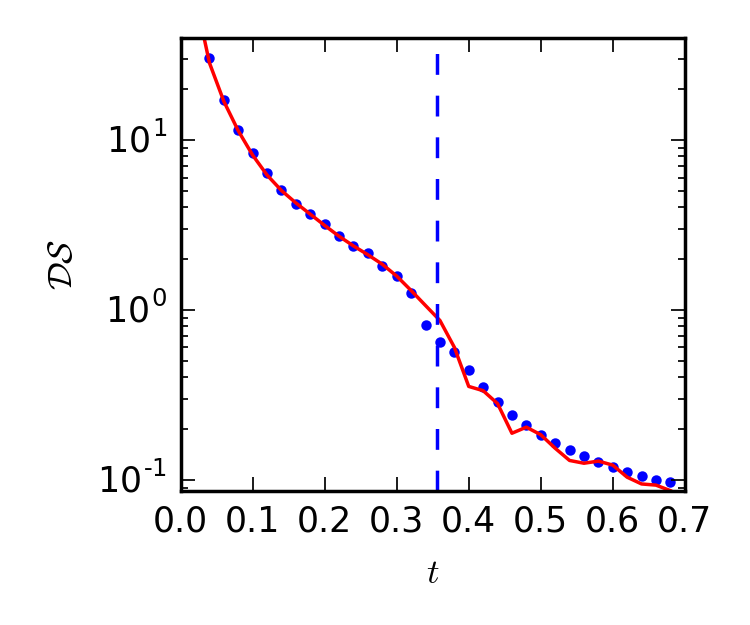}
\caption{Entanglement entropy (with $t=0$ contribution subtracted, left) and its logarithmic time derivative (right) after a quench from $\mu=10$ to $\mu=0.5$, as a function of time. The polar cap has a radius of $0.36 \text{ rad}$. The blue points correspond to the results on the fuzzy sphere with $N=400$ and the red line is an interpolation of the results on the commutative sphere with $N=199$. $N$ needs to be so high on the fuzzy sphere to accurately probe a small polar cap. The dashed vertical line indicates the expected saturation time, $t=0.36$.}
\label{f1}
\end{figure}

We start by calculating the entanglement entropy as a function of time after the quench. To get rid of the scale dependence $\mathfrak{s}_\Sigma(R)$, we also consider the logarithmic derivative of the time-dependent part of the entanglement entropy:
\begin{equation}
\mathcal{DS} \equiv \frac{1}{S(t,\theta) - S(0, \theta)} \frac{\partial}{\partial t} (S(t, \theta) - S(0, \theta))~.
\end{equation}
By construction, this quantity only depends on $F_\Sigma(t)$.
Figure \ref{f1} shows results for both the commutative and noncommutative spheres for a polar cap at a small angle. Crucially, note that the curve reaches an approximate plateau after a time equal to the angular size of the polar cap. If the entanglement spread faster on the noncommutative sphere, we would expect that feature to occur earlier. We note a small discrepancy for times close to this saturation time, but this is not a large effect. Finally, there appears to be some subleading growth after the saturation time in both the commutative and noncommutative theory. A similar effect was noted in \cite{Cotler:2016acd}.

\begin{figure}[h]
\includegraphics{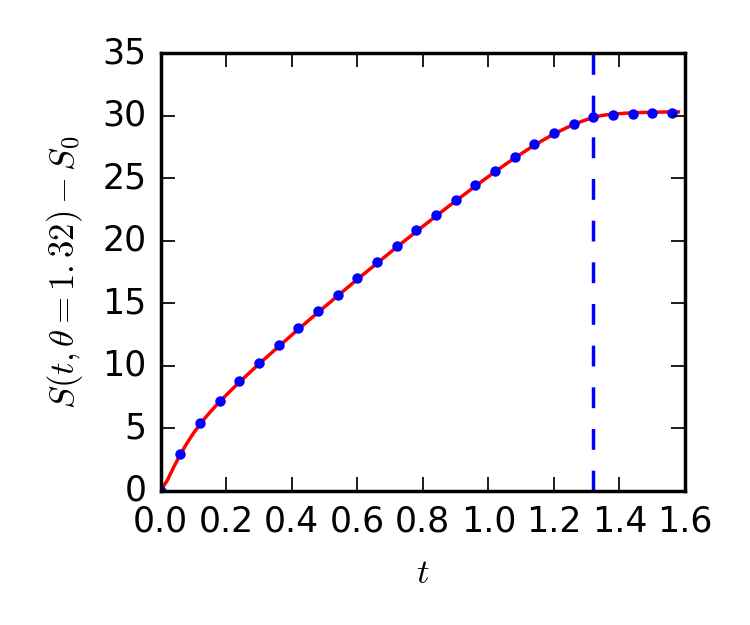}
\includegraphics{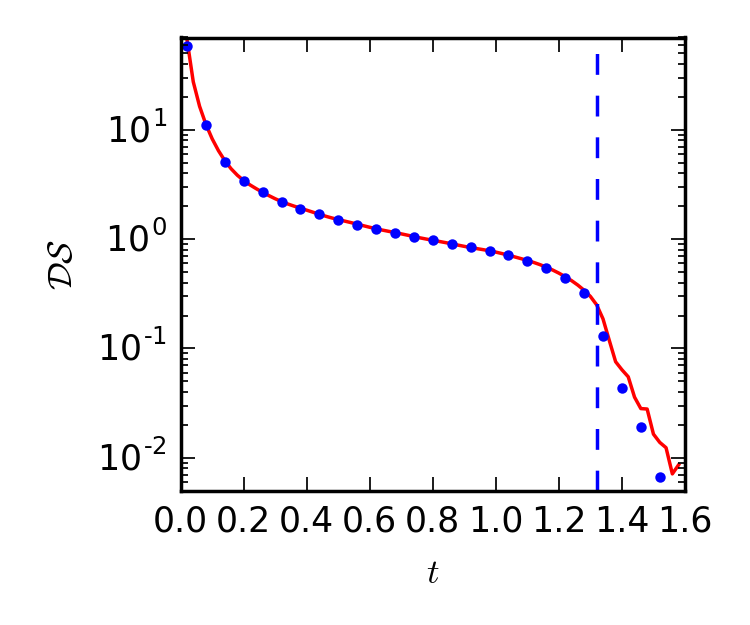}
\caption{Entanglement entropy (with $t=0$ contribution subtracted, left) and its logarithmic time derivative (right) after a quench from $\mu=10$ to $\mu=0.5$, as a function of time. The polar cap has a radius of $1.32 \text{ rad}$. The blue points correspond to the results on the fuzzy sphere with $N=200$ and the red line is an interpolation of the results on the commutative sphere with $N=213$. The dashed vertical line indicates the expected saturation time, $t=1.32$.}
\label{f2}
\end{figure}

Figure \ref{f2} shows the results for a larger polar cap. Again, we note agreement between the commutative and noncommutative results. In fact, it generally appears that for times less than but close to the expected saturation time the results for larger polar caps match the commutative result more than they do in the case of a smaller polar cap. This is probably due to edge effects.  

\begin{figure}[h!]
\includegraphics{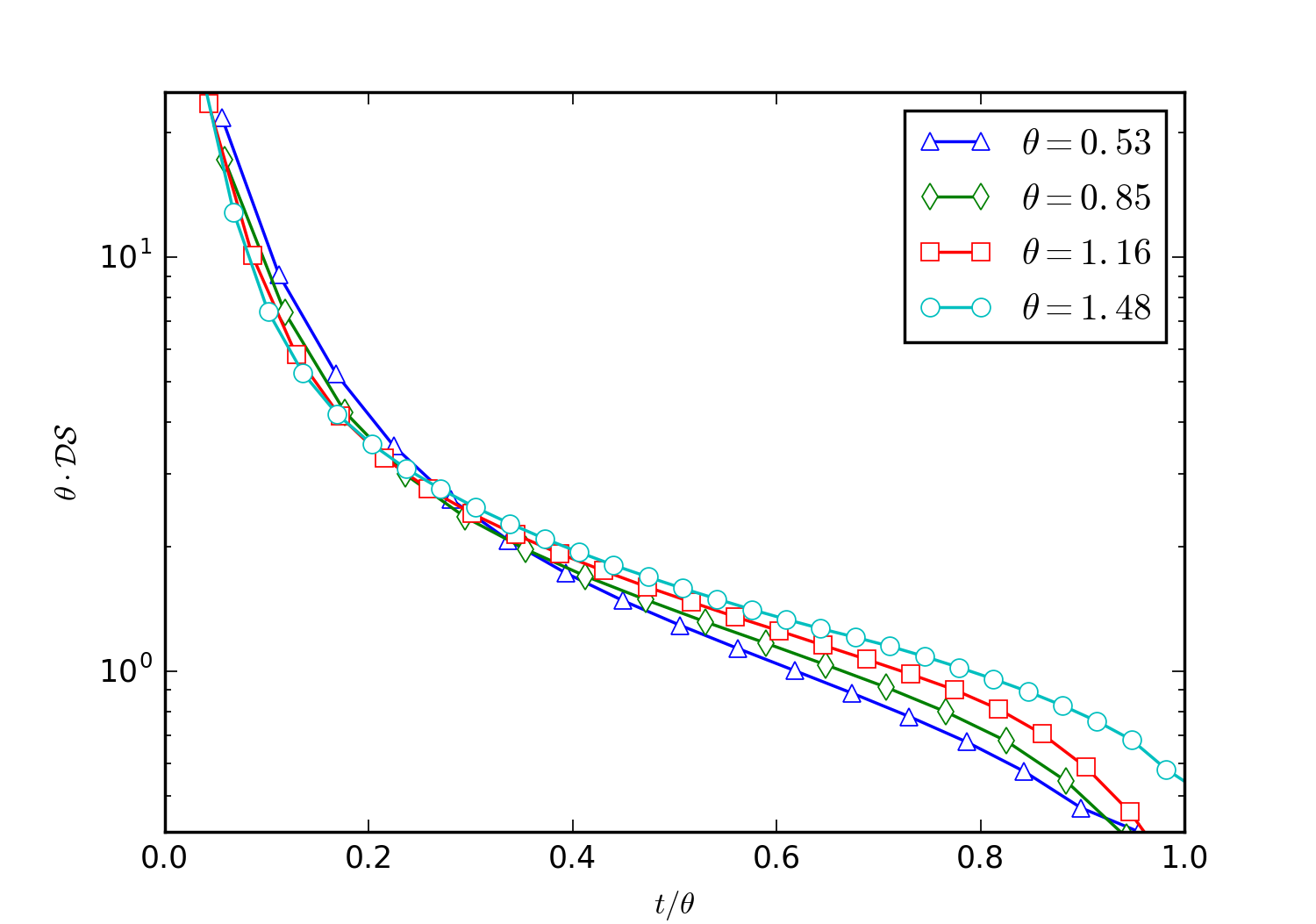}
\caption{Scaled logarithmic derivative of entanglement entropy following a mass quench as a function of time in units of the expected saturation time; for various polar cap sizes. All curves are for the theory on the commutative sphere with $N=150$ and a pre-quench mass of $\mu=10$. We can see the dependence of the global behaviour $F_\Sigma(t)$ on the size of the polar cap $\Sigma$ from the fact that the curves appear flatter as the angular size increases. Note that this effect becomes more pronounced as the time approaches the saturation time.}
\label{f3}
\end{figure}

One might expect that the function $F_\Sigma(t)$ would only depend on the ratio $t/\theta$. Indeed, if $\Sigma$ is a disk on a plane then that is the case (as in equation \ref{flat}). However, this turns out not to be the case, even in the commutative theory. Figure \ref{f3} shows $\theta \cdot \mathcal{DS}$ plotted against $t/\theta$ for various $\theta$. If $F_\Sigma(t)$ only depended on $t/\theta$ then the points for various $\theta$ would lie on the same curve in this figure, but they do not. This is not so surprising since there is another scale in the problem: the radius of the sphere.

\subsection{Mass dependence and entanglement density}
\begin{figure}[h!]
\includegraphics[scale=0.9]{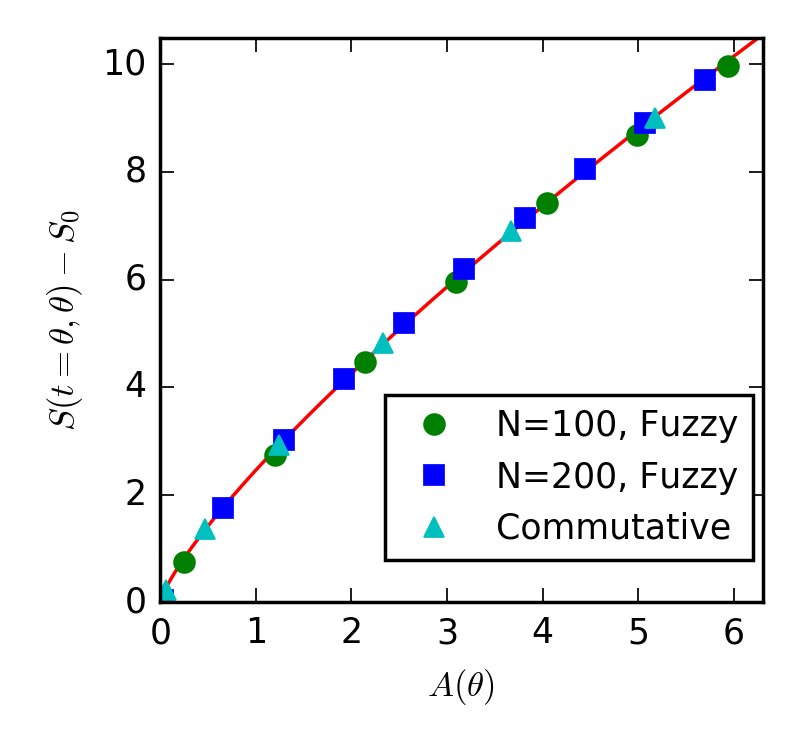}
\includegraphics[scale=0.9]{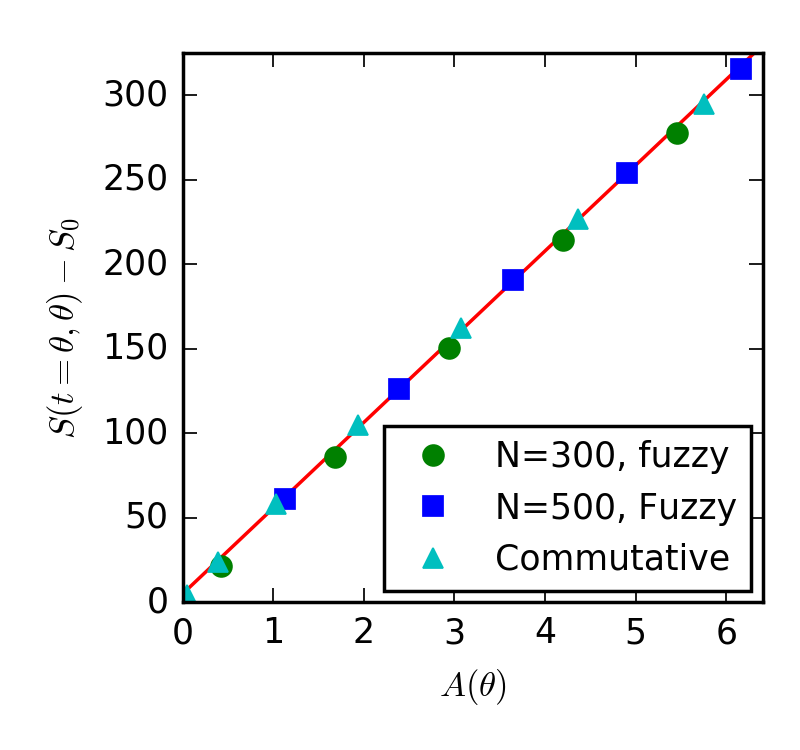}
\caption{Entanglement entropy at saturation time (with $t=0$ contribution subtracted) as a function of area. The points correspond to the result on the fuzzy sphere at different $N$ and the results on the commutative sphere, which are consistent with each other. The graph on the left is for $\mu=5$. The solid curve corresponds to a power law fit, yielding a power of 0.79. The graph on the right is for $\mu=30$. There, the solid curve is from a linear fit to the commutative results, with slope 50.74.}
\label{f4}
\end{figure}

So far, we have focused on the function $F_\Sigma(t)$ and taken the pre-quench mass to be much smaller than the UV cutoff in both theories. We now turn to how the overall scale $\mathfrak{s}_\Sigma(R)$ behaves. According to the quasi-particle model, it should be extensive:
\begin{equation}
\mathfrak{s}(\theta)=s A(\theta)~,
\end{equation}
where $A(\theta)$ is the area of the polar cap and $s$ can be thought of as an entanglement density (e.g. the number of EPR pairs generated by the quench per unit area). We find that for low pre-quench masses\footnote{Recall that $\mu$ is measured in units of the sphere radius, not of the UV cutoff.} the entanglement entropy is sub-extensive: it grows as $A^p(\theta)$ for some power $p<1$, but $p$ approaches $1$ as the mass is increased. This is independent of $N$, as long as we take care to always be in the regime \ref{smallregime}. This behaviour is seen in both the commutative and non-commutative theories and its interpretation appears simple: the temperature of the quench (which is related to the pre-quench mass) must be high enough to ``wash out'' any vacuum features. This was also discussed in \cite{Suzuki:2016sca}. These results are shown in figure \ref{f4}, which show the growth of entanglement entropy for small and large $\mu$.

We can also ask how the entanglement density depends on the pre-quench mass. In \cite{Cotler:2016acd}, it was found that on a two-dimensional plane the entanglement density for a free field quenched from mass $M$ to zero was:
\begin{equation}
s=\frac{\log 2}{4\pi} M^2~. 
\label{s-val}
\end{equation} 
The $\Phi^2$ term of the Hamiltonian of a scalar field can be written as
\begin{equation}
(M^2+\xi \mathcal{R})\Phi^2~,
\end{equation}
where $M$ is the mass of the field, $\mathcal{R}$ is the Ricci scalar of the geometry and $\xi$ is the coupling between the geometry and the field. For a unit sphere, we have that $\mathcal{R}=2$. A conformally coupled scalar in two spatial dimensions has $\xi=\frac{1}{8}$, so we can think of our parameter $\mu$ as:
\begin{equation}
\mu^2=M^2+\left(\frac{1}{2}\right)^2~,
\end{equation}
in which case going from $\mu=\mu_0$ to $\mu=0.5$ is equivalent to taking the mass of a conformally coupled scalar from $M=\sqrt{\mu_0^2-0.5^2}$ to $M=0$. From the slope of the best fit line in Figure \ref{f4}, we can see that if $s$ is proportional to $M^2$, then the proportionality constant would fall within less than 3\% of $\frac{\log 2}{4\pi}$.

\begin{figure}[h!]
\includegraphics[scale=0.9]{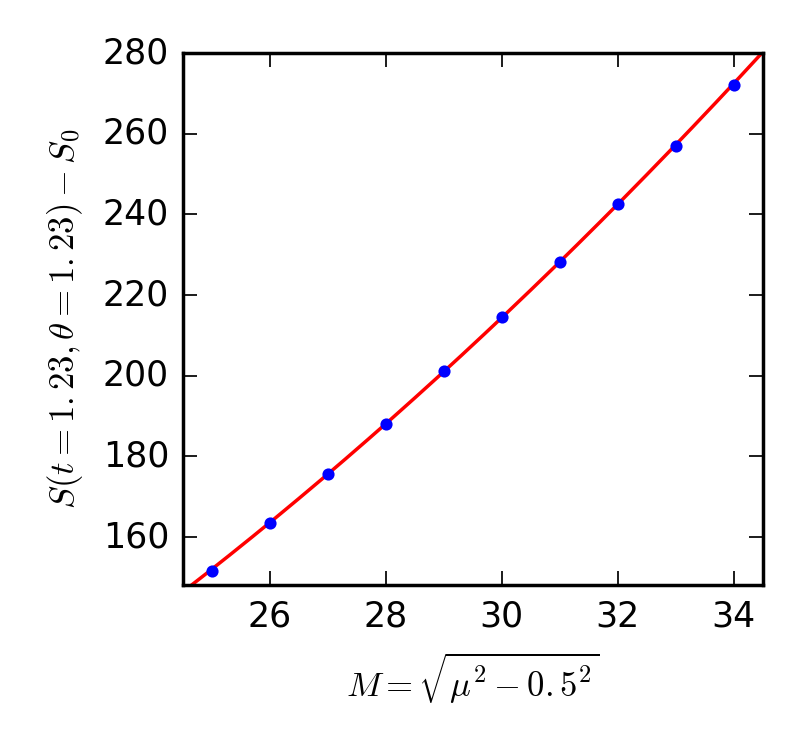}
\includegraphics[scale=0.9]{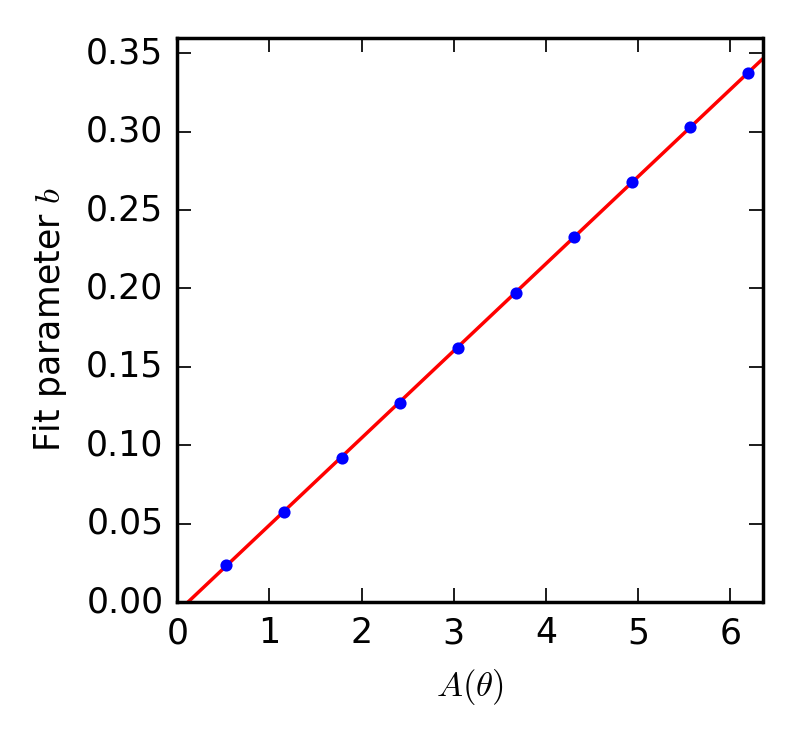}
\caption{Mass dependence of entanglement entropy at saturation time (with the $t=0$ value subtracted). The left graph is for a polar cap of size $\theta=1.23~\text{rad}$. The solid line is a quadratic fit with coefficient $b=0.227$ (and a constant term). The graph on the right shows the value of the fit parameter $b$ for polar caps of various areas (with masses taken in the same range). The slope of the fit line is 0.0555.}
\label{f5}
\end{figure}

In figure \ref{f5}, we confirm that the result \ref{s-val} applies on the fuzzy sphere. First, we consider the entanglement entropy of a polar cap of a given size at the saturation time (subtracting the $t=0$ contribution). As expected, this is quadratic. Let $b$ be the fit coefficient of $M^2$ (there is also a constant offset that accounts for the non-thermal behaviour for small $M$). We then consider how $b$ depends on the area of the polar cap: the relationship is linear, consistent with our previous observations that the entanglement entropy added by the quench is extensive. The slope of the $b$ vs $A(\theta)$ line is then the coefficient of $M^2$ in $s$. We find a value of 0.0555, which is within 0.7\% of the flat space result.

\subsection{Large masses}
We have been careful to take values of $\mu$ which are much smaller than $N$. In the case of the commutative sphere, operating in this regime is natural and we should not expect results calculated outside of it to be meaningful: any result with $\mu \sim N$ would not be physical as we take the continuum limit. However, when dealing with the fuzzy sphere the situation is more subtle. Finite $N$ is not a computational tool, it is inherent to the theory's noncommutativity. On a sphere of finite radius, it is not possible to take the UV cutoff to zero while maintaining a finite noncommutativity lengthscale. In fact, this connection between the noncommutativity scale, the UV scale and the IR scale is probably at the root of the unusual behaviour of entanglement entropy on the fuzzy sphere. 

\begin{figure}[ht!]
\includegraphics[scale=0.9]{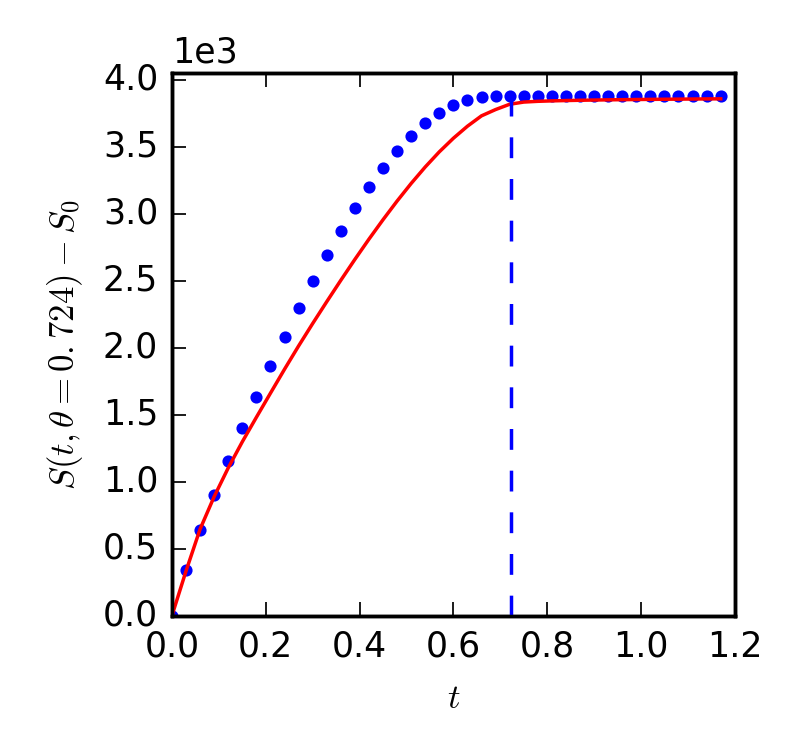}
\includegraphics[scale=0.9]{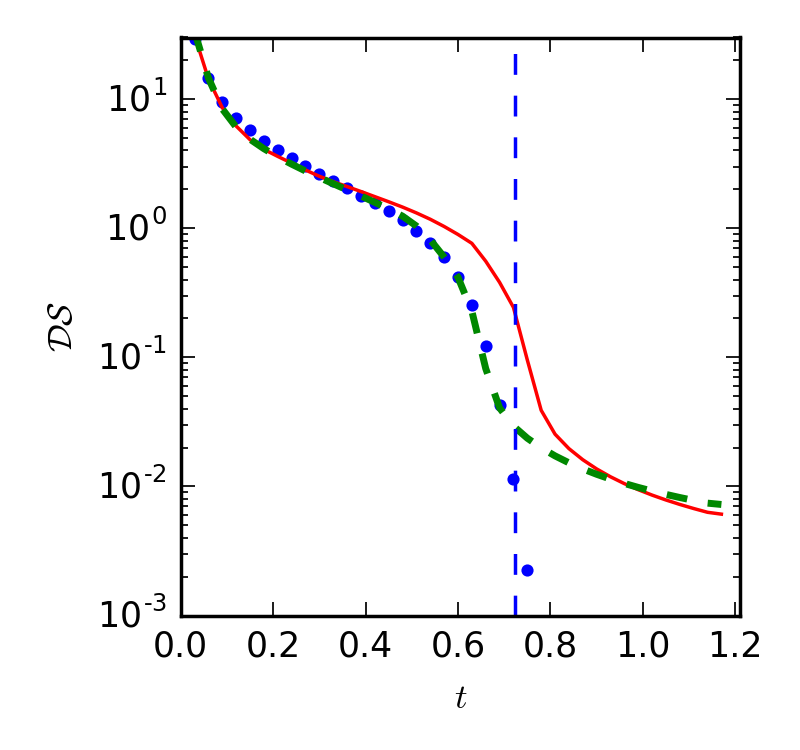}
\caption{Entanglement entropy (with $t=0$ contribution subtracted, left) and its logarithmic time derivative (right) after a quench as a function of time for a polar cap of size $0.724 \text{ rad}$. On the left, the blue points correspond to $N=200$ with $\mu=2000$ and the red line is an interpolation of the results for $\mu=20$ (scaled for comparison). On the right, the blue points correspond to $N=200$ with $\mu=2000$, the red line is an interpolation of the results for $\mu=20$. For comparison, the green dashed curve is an interpolation of results for $\mu=20$, $N=203$ for a polar cap of size $0.623 \text{ rad}$. The vertical blue dashed line corresponds to $t=0.724$.}
\label{f6}
\end{figure}

\begin{figure}[ht!]
\centering
\includegraphics[scale=0.9]{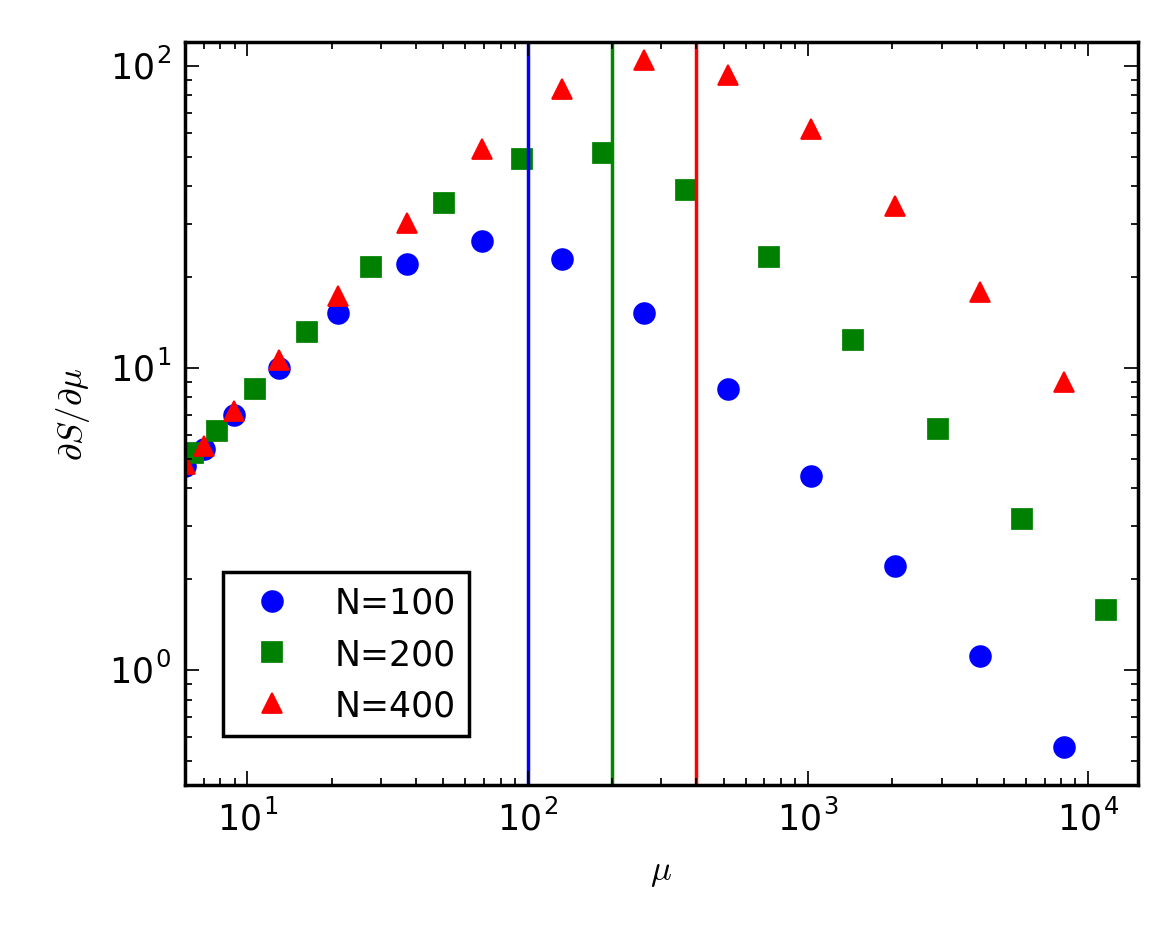}
\caption{Derivative of entanglement entropy with respect to $\mu$, evaluated at $\theta=\frac{\pi}{2}$, $t=\frac{\pi}{2}$. The various points represent different $N$, with vertical lines at those values to illustrate the turnover at $\mu \sim N$.}
\label{f7}
\end{figure}

As we move to away from the small mass regime, we find that the entanglement entropy appears to saturate faster. In fact, as we take the mass to be much larger than $N$, we find that the entanglement entropy seems to be described better by the curve describing a polar cap of reduced size, with the change in size scaling roughly as $1/\sqrt{N}$. An example of this is shown in Figure \ref{f6}. It is interesting to note that this is compatible with the idea that the theory is non-local on scales up to the noncommutativity lengthscale. To see that this change in behaviour occurs at $\mu \sim N$, we can consider the derivative of the entanglement entropy of a region at the saturation time as a function of $\mu$ for various $N$. This is shown on a log-log scale in figure \ref{f7}. Notice that for small $\mu$ we have a line of slope 1, which is expected for the $M^2$ growth described earlier. For large $\mu$, we have a line of slope -1, indicating logarithmic growth. The position of the turnover point is clearly close to $N$. Note that a qualitative change in the behaviour of the entanglement entropy at large masses was also seen at $t=0$ in \cite{Karczmarek:2013jca}.

\section*{Acknowledgements}
I am grateful to Melissa Anholm, Joanna Karczmarek, Napat Poovuttikul, Aurelio Romero-Berm\'{u}dez, Koenraad Schalm and Jan Zaanen for valuable discussions. This work is part of the research programme of the Foundation for Fundamental Research on Matter (FOM), which is part of the Netherlands Organisation for Scientific Research (NWO).
\clearpage
\bibliographystyle{JHEP}
\bibliography{thebib}

\providecommand{\href}[2]{#2}\begingroup\raggedright\begin{thebibliography}{10}

\bibitem{Ryu:2006bv}
S.~Ryu and T.~Takayanagi, \emph{{Holographic derivation of entanglement entropy
  from AdS/CFT}},
  \href{http://dx.doi.org/10.1103/PhysRevLett.96.181602}{\emph{Phys. Rev.
  Lett.} {\bfseries 96} (2006) 181602},
  [\href{https://arxiv.org/abs/hep-th/0603001}{{\ttfamily hep-th/0603001}}].

\bibitem{Hubeny:2007xt}
V.~E. Hubeny, M.~Rangamani and T.~Takayanagi, \emph{{A Covariant holographic
  entanglement entropy proposal}},
  \href{http://dx.doi.org/10.1088/1126-6708/2007/07/062}{\emph{JHEP} {\bfseries
  07} (2007) 062}, [\href{https://arxiv.org/abs/0705.0016}{{\ttfamily
  0705.0016}}].

\bibitem{Liu:2013iza}
H.~Liu and S.~J. Suh, \emph{{Entanglement Tsunami: Universal Scaling in
  Holographic Thermalization}},
  \href{http://dx.doi.org/10.1103/PhysRevLett.112.011601}{\emph{Phys. Rev.
  Lett.} {\bfseries 112} (2014) 011601},
  [\href{https://arxiv.org/abs/1305.7244}{{\ttfamily 1305.7244}}].

\bibitem{Calabrese:2005in}
P.~Calabrese and J.~L. Cardy, \emph{{Evolution of entanglement entropy in
  one-dimensional systems}},
  \href{http://dx.doi.org/10.1088/1742-5468/2005/04/P04010}{\emph{J. Stat.
  Mech.} {\bfseries 0504} (2005) P04010},
  [\href{https://arxiv.org/abs/cond-mat/0503393}{{\ttfamily
  cond-mat/0503393}}].

\bibitem{Kundu:2016cgh}
S.~Kundu and J.~F. Pedraza, \emph{{Spread of entanglement for small subsystems
  in holographic CFTs}},
  \href{http://dx.doi.org/10.1103/PhysRevD.95.086008}{\emph{Phys. Rev.}
  {\bfseries D95} (2017) 086008},
  [\href{https://arxiv.org/abs/1602.05934}{{\ttfamily 1602.05934}}].

\bibitem{Casini:2015zua}
H.~Casini, H.~Liu and M.~Mezei, \emph{{Spread of entanglement and causality}},
  \href{http://dx.doi.org/10.1007/JHEP07(2016)077}{\emph{JHEP} {\bfseries 07}
  (2016) 077}, [\href{https://arxiv.org/abs/1509.05044}{{\ttfamily
  1509.05044}}].

\bibitem{Cotler:2016acd}
J.~S. Cotler, M.~P. Hertzberg, M.~Mezei and M.~T. Mueller, \emph{{Entanglement
  Growth after a Global Quench in Free Scalar Field Theory}},
  \href{http://dx.doi.org/10.1007/JHEP11(2016)166}{\emph{JHEP} {\bfseries 11}
  (2016) 166}, [\href{https://arxiv.org/abs/1609.00872}{{\ttfamily
  1609.00872}}].

\bibitem{Seiberg:1999vs}
N.~Seiberg and E.~Witten, \emph{{String theory and noncommutative geometry}},
  \href{http://dx.doi.org/10.1088/1126-6708/1999/09/032}{\emph{JHEP} {\bfseries
  09} (1999) 032}, [\href{https://arxiv.org/abs/hep-th/9908142}{{\ttfamily
  hep-th/9908142}}].

\bibitem{Bigatti:1999iz}
D.~Bigatti and L.~Susskind, \emph{{Magnetic fields, branes and noncommutative
  geometry}}, \href{http://dx.doi.org/10.1103/PhysRevD.62.066004}{\emph{Phys.
  Rev.} {\bfseries D62} (2000) 066004},
  [\href{https://arxiv.org/abs/hep-th/9908056}{{\ttfamily hep-th/9908056}}].

\bibitem{Sekino:2008he}
Y.~Sekino and L.~Susskind, \emph{{Fast Scramblers}},
  \href{http://dx.doi.org/10.1088/1126-6708/2008/10/065}{\emph{JHEP} {\bfseries
  10} (2008) 065}, [\href{https://arxiv.org/abs/0808.2096}{{\ttfamily
  0808.2096}}].

\bibitem{Lashkari:2011yi}
N.~Lashkari, D.~Stanford, M.~Hastings, T.~Osborne and P.~Hayden, \emph{{Towards
  the Fast Scrambling Conjecture}},
  \href{http://dx.doi.org/10.1007/JHEP04(2013)022}{\emph{JHEP} {\bfseries 04}
  (2013) 022}, [\href{https://arxiv.org/abs/1111.6580}{{\ttfamily 1111.6580}}].

\bibitem{Brady:2013opa}
L.~Brady and V.~Sahakian, \emph{{Scrambling with Matrix Black Holes}},
  \href{http://dx.doi.org/10.1103/PhysRevD.88.046003}{\emph{Phys. Rev.}
  {\bfseries D88} (2013) 046003},
  [\href{https://arxiv.org/abs/1306.5200}{{\ttfamily 1306.5200}}].

\bibitem{Karczmarek:2013jca}
J.~L. Karczmarek and P.~Sabella-Garnier, \emph{{Entanglement entropy on the
  fuzzy sphere}}, \href{http://dx.doi.org/10.1007/JHEP03(2014)129}{\emph{JHEP}
  {\bfseries 03} (2014) 129},
  [\href{https://arxiv.org/abs/1310.8345}{{\ttfamily 1310.8345}}].

\bibitem{Sabella-Garnier:2014fda}
P.~Sabella-Garnier, \emph{{Mutual information on the fuzzy sphere}},
  \href{http://dx.doi.org/10.1007/JHEP02(2015)063}{\emph{JHEP} {\bfseries 02}
  (2015) 063}, [\href{https://arxiv.org/abs/1409.7069}{{\ttfamily 1409.7069}}].

\bibitem{Okuno:2015kuc}
S.~Okuno, M.~Suzuki and A.~Tsuchiya, \emph{{Entanglement entropy in scalar
  field theory on the fuzzy sphere}},
  \href{http://dx.doi.org/10.1093/ptep/ptv192}{\emph{PTEP} {\bfseries 2016}
  (2016) 023B03}, [\href{https://arxiv.org/abs/1512.06484}{{\ttfamily
  1512.06484}}].

\bibitem{Suzuki:2016sca}
M.~Suzuki and A.~Tsuchiya, \emph{{A generalized volume law for entanglement
  entropy on the fuzzy sphere}},
  \href{https://arxiv.org/abs/1611.06336}{{\ttfamily 1611.06336}}.

\bibitem{Fischler:2013gsa}
W.~Fischler, A.~Kundu and S.~Kundu, \emph{{Holographic Entanglement in a
  Noncommutative Gauge Theory}},
  \href{http://dx.doi.org/10.1007/JHEP01(2014)137}{\emph{JHEP} {\bfseries 01}
  (2014) 137}, [\href{https://arxiv.org/abs/1307.2932}{{\ttfamily 1307.2932}}].

\bibitem{Karczmarek:2013xxa}
J.~L. Karczmarek and C.~Rabideau, \emph{{Holographic entanglement entropy in
  nonlocal theories}},
  \href{http://dx.doi.org/10.1007/JHEP10(2013)078}{\emph{JHEP} {\bfseries 10}
  (2013) 078}, [\href{https://arxiv.org/abs/1307.3517}{{\ttfamily 1307.3517}}].

\bibitem{Shiba:2013jja}
N.~Shiba and T.~Takayanagi, \emph{{Volume Law for the Entanglement Entropy in
  Non-local QFTs}},
  \href{http://dx.doi.org/10.1007/JHEP02(2014)033}{\emph{JHEP} {\bfseries 02}
  (2014) 033}, [\href{https://arxiv.org/abs/1311.1643}{{\ttfamily 1311.1643}}].

\bibitem{Pang:2014tpa}
D.-W. Pang, \emph{{Holographic entanglement entropy of nonlocal field
  theories}}, \href{http://dx.doi.org/10.1103/PhysRevD.89.126005}{\emph{Phys.
  Rev.} {\bfseries D89} (2014) 126005},
  [\href{https://arxiv.org/abs/1404.5419}{{\ttfamily 1404.5419}}].

\bibitem{MohammadiMozaffar:2017nri}
M.~R. Mohammadi~Mozaffar and A.~Mollabashi, \emph{{Entanglement in
  Lifshitz-type Quantum Field Theories}},
  \href{https://arxiv.org/abs/1705.00483}{{\ttfamily 1705.00483}}.

\bibitem{Madore:1991bw}
J.~Madore, \emph{{The Fuzzy sphere}},
  \href{http://dx.doi.org/10.1088/0264-9381/9/1/008}{\emph{Class. Quant. Grav.}
  {\bfseries 9} (1992) 69--88}.

\bibitem{Douglas:2001ba}
M.~R. Douglas and N.~A. Nekrasov, \emph{{Noncommutative field theory}},
  \href{http://dx.doi.org/10.1103/RevModPhys.73.977}{\emph{Rev. Mod. Phys.}
  {\bfseries 73} (2001) 977--1029},
  [\href{https://arxiv.org/abs/hep-th/0106048}{{\ttfamily hep-th/0106048}}].

\bibitem{Dou:2006ni}
D.~Dou and B.~Ydri, \emph{{Entanglement entropy on fuzzy spaces}},
  \href{http://dx.doi.org/10.1103/PhysRevD.74.044014}{\emph{Phys. Rev.}
  {\bfseries D74} (2006) 044014},
  [\href{https://arxiv.org/abs/gr-qc/0605003}{{\ttfamily gr-qc/0605003}}].

\bibitem{Srednicki:1993im}
M.~Srednicki, \emph{{Entropy and area}},
  \href{http://dx.doi.org/10.1103/PhysRevLett.71.666}{\emph{Phys. Rev. Lett.}
  {\bfseries 71} (1993) 666--669},
  [\href{https://arxiv.org/abs/hep-th/9303048}{{\ttfamily hep-th/9303048}}].

\bibitem{Casini:2009sr}
H.~Casini and M.~Huerta, \emph{{Entanglement entropy in free quantum field
  theory}}, \href{http://dx.doi.org/10.1088/1751-8113/42/50/504007}{\emph{J.
  Phys.} {\bfseries A42} (2009) 504007},
  [\href{https://arxiv.org/abs/0905.2562}{{\ttfamily 0905.2562}}].

\end{thebibliography}\endgroup

\end{document}